\documentclass[12pt]{iopart}
\usepackage{graphicx}
  
\begin{document}

\title[Anomalous Infiltration]{Anomalous Infiltration}

\author{Nickolay Korabel and Eli Barkai}

\address{Physics Department, Institute of Nanotechnology and Advanced Materials, Bar-Ilan University, Ramat-Gan 52900, Israel}
\ead{nnkorabel@gmail.com, barkaie@mail.biu.ac.il}
\begin{abstract}
Infiltration of anomalously diffusing particles from one material to another through a biased interface is studied using continuous time random walk and L\'evy walk approaches. Subdiffusion in both systems may lead to a net drift from one material to another (e.g. $\left< x(t)\right> >0$) even if particles eventually flow in the opposite direction (e.g. number of particles in $x>0$ approaches zero). A weaker paradox is found for a symmetric interface: a flow of particles is observed while the net drift is zero. For a subdiffusive sample coupled to a superdiffusive system we calculate the average occupation fractions and the scaling of the particles distribution. We find a net drift in this system, which is always directed to the superdiffusive material, while the particles flow to the material with smaller sub or superdiffusion exponent. We report the exponents of the first passage times distribution of L\'evy walks, which are needed for the calculation of anomalous infiltration.

\end{abstract}

\pacs{02.50.-r, 05.40.Fb, 05.10.Gg}
\vspace{2pc}
\noindent{\it Keywords}: infiltration, anomalous diffusion, continuous time random walk, L\'evy walk
\maketitle

\section{Introduction}\label{Intro}

Infiltration of diffusing particles from material A to material B through some interface is a widely investigated process. In recent years much focus was diverted to the problems when the diffusion in one material or in both is anomalous, namely $\left< x^2(t) \right> \propto t^{\alpha}$ and $\alpha \ne 1$ \cite{BG,MK,KS05}. This behavior is important in numerous applications such as infiltration of water into porous soil \cite{Abd,Klemm}, contaminant diffusion \cite{Kirchner,Fomin}, moisture ingress in zeolites \cite{Azevedo} or in fired clay ceramics \cite{Kuntz,Wilson}, diffusion of sugar through a membrane in a gel solvent \cite{Kos}, and polymer translocation through a nanopore \cite{KK,Luo,Wanunu} (see also \cite{Jacobson,Kusumi,Morph}). 

We consider two semi-infinite materials located in $x<0$ and $x>0$, where particles exhibit anomalous  sub or superdiffusion. We find several peculiar behaviors unique to anomalous infiltration: in the case of a composition of two subdiffusive systems the infiltration of particles from material $x<0$ to $x>0$ leads to a net drift $\left< x(t) \right>$. Such a drift increases slower than $t^{1/2}$ as expected from unbiased diffusion processes, still it may yield a net sub-current $j \simeq d\left< x \right>/dt$, which vanishes as $t \rightarrow \infty$. We also show that in some cases the flow of particles is opposite to the drift. In fact we find a situation when asymptotically all the particles are say in sample $x>0$ but the average drift $\left< x(t) \right>$ is oppositely directed  $\left< x(t) \right><0$. This seemingly paradoxical behavior is explained in the text. Secondly, if materials $x<0$ and $x>0$ have different diffusive properties, in the long time limit all particles will be accumulated in the material with slower diffusion, which will act as a trap producing a flow of particles from one material to another. Interestingly, in the long time limit the drift depends only on the properties of the slower medium. This is a surprising result since $\left< x(t) \right>$ can be very far from the interface, deep in the faster sample, but still be independent of the properties of that region.

A second model we study is a subdiffusive material (for example in $x<0$) coupled to superdiffusive sample (in $x>0$). For superdiffusion motion we consider a L\'evy walk model \cite{SWK,ZK93,ZKB93}. To analyze this infiltration problem we need the distribution of the first passage times (FPT) \cite{Redner} for a L\'evy walk on a semi-axes, which are reported here for the first time (see \cite{Barkai01,Lind,Ben,Cond} for other works on anomalous first passage time problem). Using the FPT density in $x>0$ and $x<0$, we calculate the average of occupation fractions and find the scaling of the particles distribution. For a subdiffusive system coupled to a superdiffusive material a net drift is found even for unbiased motion on the boundary. This drift is always directed to the superdiffusive material, while the particles flow to the material with longer sticking times. 

Although phenomena such as drift against the flow and flow without the drift are known for systems with normal diffusion where they are generated by geometrical constraints, or by thermal or external field inhomogeneities \cite{Collins,Ostrowsky,Milligen,Burdzy}, in our case these effects are only due to the anomalous nature of diffusion and are not present for normal diffusion. These phenomena are explained by the competition of the diffusion processes which are slower or faster than normal spreading. Part of the results were shortly summarized in \cite{KB2010}. 

\section{Two Coupled Subdiffusive Systems}

We consider two coupled subdiffusive materials using the continuous time random walk (CTRW) model as the underlying process \cite{CTRW}. In the continuum limit the fractional diffusion equations in materials $x<0$ and $x>0$ describe the dynamics \cite{MK,SW,Balakrishnan,MBK99,MBK00}
\begin{eqnarray}
\label{anomaleq}
\frac{\partial P(x,t)}{\partial t} = \; _{0}D_{t}^{1-\alpha^{-}} K^{-} \frac{\partial^2}{\partial x^2}P(x,t), \quad x<0, \nonumber \\
\frac{\partial P(x,t)}{\partial t} = \; _{0}D_{t}^{1-\alpha^{+}} K^{+} \frac{\partial^2}{\partial x^2}P(x,t), \quad x>0,
\end{eqnarray}
where the Riemann-Liouville operator $_{0}D_{t}^{1-\alpha}$ is defined as \cite{MK,Podlubny}
\begin{equation}
\label{RLoperator}
_{0}D_{t}^{1-\alpha} P(x,t) = \frac{1}{\Gamma(\alpha)} \frac{\partial}{\partial t} \int_{0}^{t} dt' \frac{P(x,t')}{(t-t')^{1-\alpha}},
\end{equation}
and $0 < \alpha^{-} \le 1$, $0 < \alpha^{+} \le 1$. Constants $K^{-}$, $K^{+}$ are anomalous diffusion coefficients with units $[m^2/sec^{\alpha^{-}}]$ and $[m^2/sec^{\alpha^{+}}]$, respectively. As well known, the fractional diffusion equation (\ref{anomaleq}) with $\alpha^{-}=\alpha^{+}=\alpha$ and $K^{-}=K^{+}=K$ yields for particles starting on the origin $\left< x^2(t) \right> = 2 K t^{\alpha}/\Gamma (1+\alpha)$ \cite{MK,SW,Balakrishnan,MBK99,MBK00}. To solve equation (\ref{anomaleq}) we determine the boundary conditions for this equation starting with a random walk picture.

\subsection{CTRW model: Drift}

\begin{figure}[t!]
  \centering
  \includegraphics[width=.5\textwidth]{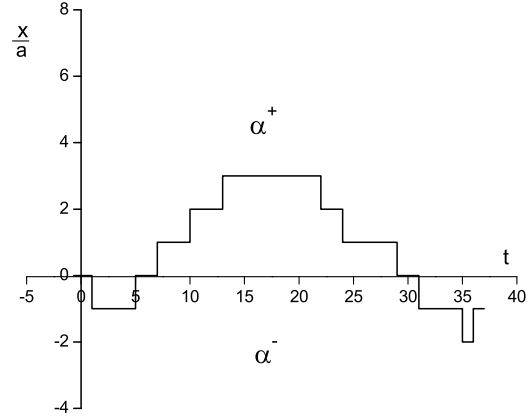}
  \caption{Illustration of the random walk model. The times which particles spend at sites in the region $x<0$ and $x>0$ are distributed by the waiting time PDFs $\psi^{-}$ and $\psi^{+}$, respectively. The waiting times at $x=0$ are exponentially distributed.}
  \label{FIG_RW}
\end{figure}

We consider a CTRW on a one dimensional lattice (see figure \ref{FIG_RW}) with the lattice spacing $a$, which in the continuum limit will be made small. For lattice points $x<0$ a particle has the probability $1/2$ to jump to one of its nearest neighbors. Waiting times on each lattice point are independent identically distributed random variables with a common PDF $\psi^{-}(\tau)$. For $x>0$ a similar unbiased random walk takes place with a waiting time PDF $\psi^{+}(\tau)$. On the lattice point $x=0$ (the boundary) a particle has the probability to jump right $q^{+}$ or left $q^{-}=1-q^{+}$ and the waiting times are exponentially distributed with a rate $R_0$. Thus, a particle starting on the origin will jump say to the right (with probability $q^{+}$) after waiting an average time $1/R_0$, then on the lattice point $x=a$, it will wait for time $\tau$ drawn from $\psi^{+}(\tau)$, and then with probability $1/2$ will jump to the left or right. The waiting times have power law distributions $\psi^{-}(\tau) \propto \tau^{-1-\alpha^{-}}$ and $\psi^{+}(\tau) \propto \tau^{-1-\alpha^{+}}$, as $\tau \rightarrow \infty$. More specifically, using Tauberian theorem the Laplace transform $\tau \rightarrow s$ of the waiting time PDF behaves like
\begin{equation}
\label{psi}
\tilde{\psi}^{-}(s) \propto 1 - B^{-} s^{\alpha^{-}}, \; \; \; \; \tilde{\psi}^{+}(s) \propto 1 - B^{+} s^{\alpha^{+}},
\end{equation}
when $s \rightarrow 0$ corresponding to $\tau \rightarrow \infty$. Throughout the paper we denote the Laplace transform as $\tilde{f}(s)=\int_0^{\infty} dt \; e^{-st} f(t)$. The anomalous diffusion coefficients are given by $K^{-} = \lim_{a^2 \rightarrow 0, B^{-} \rightarrow 0} a^2/ 2 B^{-}$ and $K^{+} = \lim_{a^2 \rightarrow 0, B^{+} \rightarrow 0} a^2/2 B^{+}$ \cite{MBK00}. Our results are not changed if on $x=0$ the waiting times are power law distributed like $\psi^{-}$ or $\psi^{+}$ instead of exponential.

\begin{figure}[t!]
  \centering
  \includegraphics[width=.65\textwidth]{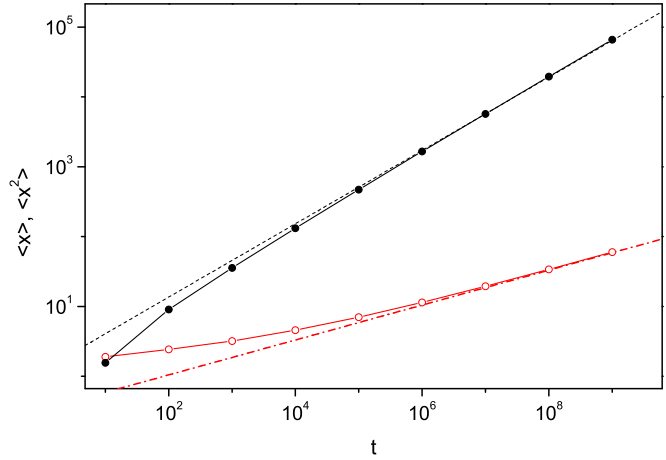}
  \caption{The drift $\left< x(t) \right>$ (open circles) and $\left< x^2 \right>$ (filled circles) calculated numerically for the CTRW model with $\alpha^{+}=0.75$, $K^{+}=0.138$, $\alpha^{-}=0.3$, $K^{-}=0.385$ and $q^{+}=0.7$. Dashed and dashed-dotted lines represent long time asymptotic behavior described by equations (\ref{xmean3}) and (\ref{x2long}), respectively.}
  \label{FIG}
\end{figure}

Using the CTRW model we find the drift in the long time limit in the following way: From the start of the process at $t=0$ until time $t$ the particle made $N$ jumps, where $N$ is a random variable. Hence the position of the particle at time $t$ is
\begin{equation}
\label{x}
x = \sum_{i=1}^{N} \delta x_i,
\end{equation}
if at $t=0$ the particle is on the origin. From the model assumptions $\delta x_i$ is equal $+a$ or $-a$ and it describes the $i$th jump in the sequence. The jump length $\delta x_i$ satisfies $\left< \delta x_i \right>=0$ if the particle is not on the origin since then the probability of jumping left or right is equal $1/2$. Therefore
\begin{equation}
\label{xmean}
\left< x(t) \right> = a (q^{+} - q^{-}) \left< n_z(t) \right>,
\end{equation}
where $\left< n_z(t) \right>$ is the average number of times the particle visited the origin, which is calculated in the Appendix A using the CTRW model. In particular $\left< n_z(t) \right>$ is determined by the first passage time PDFs in samples $x>0$ and $x<0$ since these first passage times determine the number of visits to the origin (see details in Appendix A). To derive equation (\ref{xmean}) we have used the fact that on the origin the average step size is $a (q^{+} - q^{-})$. In \cite{KB10} we related $q^{+}, q^{-}$ to energy gap between material A and B using detailed balance condition. When $\alpha^{-}=\alpha^{+}=\alpha$
\begin{equation}
\label{xmean2}
\left< x(t) \right> \propto \frac{q^{+} - q^{-}}{\Gamma(1+\alpha/2)} \; \frac{\sqrt{K^{-} K^{+}}}{q^{-} \sqrt{K^{+}} + q^{+} \sqrt{K^{-}}} \; t^{\alpha/2}.
\end{equation}
For the case $\alpha^{-}<\alpha^{+}$, we get
\begin{equation}
\label{xmean3}
\left< x(t) \right> \propto \frac{q^{+} - q^{-}}{q^{-}} \frac{\sqrt{K^{-}}}{\Gamma(1+\alpha^{-}/2)} \; t^{\alpha^{-}/2},
\end{equation}
which agrees well with simulations in figure \ref{FIG}. The sign of the drift, i.e. its directionality, is determined by the sign of $q^{+} - q^{-}$, and $\left< x(t) \right>=0$ if $q^{+}=q^{-}$. Equation (\ref{xmean3}) shows that in the long time limit the drift depends only on one diffusion constant in sample $x<0$ (i.e. $K^{-}$) and grows in time with the exponent of the slower medium (i.e. $\alpha^{-}$). This is a surprising result since $\left< x(t)\right>$ can be very far from the interface, deep in the faster sample $x>0$, but still be independent of the properties of that region $\alpha^{+}$, $K^{+}$. The exponent of the drift $\left< x(t) \right> \propto t^{\alpha^{-}/2}$ is the exponent of the slower medium, which is clearly related to the power law distribution of first passage times in the slower medium $\phi^{-}(t) \propto t^{-(1+\alpha^{-}/2)}$ \cite{Barkai01} (see Appendix A). One can show that equation (\ref{xmean3}) is valid for times
\[
1 \ll \sqrt{\frac{K^{+}}{K^{-}}} \frac{q^{-}}{q^{+}} \; t^{(\alpha^{+}-\alpha^{-})/2}.
\]

\subsection{CTRW model: Statistics of occupation times}

Now we consider the distribution of occupation times in the material $x<0$ or $x>0$. Let $f_t(t^{-})$ be the PDF of the total time $t^{-}$ a walker stays in the material $x<0$ and $t$ is the measurement time. Similarly, $t^{+}$ is the total time a walker stays in the material $x>0$. The double Laplace transform of $f_t(t^{-})$, $\tilde{f}_{s}(u)=\int_0^{\infty} dt e^{-s t} \int_0^{\infty} dt^{-} e^{-u t^{-}} f_t(t^{-})$, reads (the derivation is given in Appendix B)
\begin{equation}
\label{ofrac}
\tilde{f}_{s}(u) \approx \frac{q^{-} (s+u)^{\frac{\alpha^{-}}{2}-1}/\sqrt{K^{-}} + q^{+} s^{\frac{\alpha^{+}}{2}-1}/\sqrt{K^{+}}}{q^{-} (s+u)^{\frac{\alpha^{-}}{2}}/\sqrt{K^{-}} + q^{+} s^{\frac{\alpha^{+}}{2}}/\sqrt{K^{+}}}.
\end{equation}
For $\alpha^{-}=\alpha^{+}$ equation (\ref{ofrac}) reduces to the Lamperti PDF \cite{GL01,BB06} (see Appendix B). 

\begin{figure}[t!]
  \centering
  \includegraphics[width=.65\textwidth]{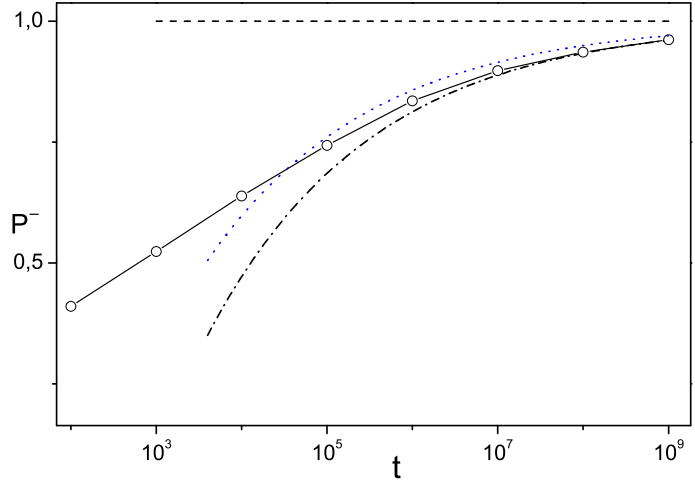}
  \caption{The occupation fraction in $x<0$, $\mathcal{P}^{-}(t)=\int_{-\infty}^{0} dx \; P(x,t)$, calculated numerically for the CTRW model (open circles). Parameters are the same as in figure \ref{FIG}. The dashed-dotted line given by equation (\ref{avfrac2}) describes how $\mathcal{P}^{-}$ approach its limit $\mathcal{P}^{-} \rightarrow 1$ (dashed line). The dotted line corresponds to $\mathcal{P}^{-}(t)$ calculated using equation (\ref{Pp}). Notice that all the particles flow to the left, however $\left< x(t) \right> >0$ namely particles drift to the right (see figure \ref{FIG}).}
  \label{FIG_P+}
\end{figure}

Let us assume $\alpha^{-}<\alpha^{+}$. Expanding equation (\ref{ofrac}) in $u$, we get
\begin{equation}
\label{t_minus}
\left< \tilde{t}^{-}(s) \right>  \propto \frac{1}{s^2(1+\tilde{\mathcal{R}}(s))},
\end{equation}
where $\tilde{\mathcal{R}}(s)$ is given by
\begin{equation}
\label{R}
\tilde{\mathcal{R}}(s) = \frac{q^{+}}{q^{-}} \frac{\sqrt{K^{-}}}{\sqrt{K^{+}}} \frac{s^{\alpha^{+}/2}}{s^{\alpha^{-}/2}}.
\end{equation}
In the long time limit $t \rightarrow \infty$ ($s \rightarrow 0$) equation (\ref{t_minus}) gives (after inverting the Laplace transform)
\begin{equation}
\left< t^{-}(t) \right>  \propto t - \frac{q^{+}}{q^{-}} \sqrt{\frac{K^{-}}{K^{+}}} \frac{t^{1-\frac{\alpha^{+}-\alpha^{-}}{2}}}{\Gamma(2-\frac{\alpha^{+}-\alpha^{-}}{2})}.
\end{equation}
Now it is straightforward to get the average occupation fraction in sample $x<0$ 
\begin{equation}
\label{avfrac2}
\mathcal{P}^{-}(t) = \frac{\left< t^{-}(t) \right>}{t} \propto 1 - \frac{q^{+}}{q^{-}} \sqrt{\frac{K^{-}}{K^{+}}} \frac{t^{\frac{\alpha^{-}-\alpha^{+}}{2}}}{\Gamma(2+\frac{\alpha^{-}-\alpha^{+}}{2})}.
\end{equation}
Since $\alpha^{-}<\alpha^{+}$, the second term in this equation vanishes as $t \rightarrow \infty$ and $\mathcal{P}^{-}(t) \rightarrow 1$, which indicates that in the long time limit all particles will be located in the region $x<0$ (see figure \ref{FIG_P+}), a result valid for any $q^{-}, q^{+}<1$. As might be expected, the slow domain $\alpha^{-}<\alpha^{+}$ serves as a perfect trap: all particles flow to the slower domain. The usual normalization condition 
\begin{equation}
\mathcal{P}^{-}(t) + \mathcal{P}^{+}(t)=1,
\end{equation} 
gives $\mathcal{P}^{+}(t)$ in sample $x>0$. 

\subsection{Boundary conditions and solution of equation (\ref{anomaleq})}

We now wish to find $P(x,t)$ i.e. solve equation (\ref{anomaleq}). To obtain the solution of a standard or fractional diffusion equation, one has to know the mathematical boundary conditions between sample $x<0$ and sample $x>0$. One boundary condition is well known and needs no further discussion: the probability current must be balanced so that normalization is conserved (conservation of number of particles), see some details below. Previous works assumed in addition the second boundary condition which demand the constant ratio of the concentrations at two sides of the boundary located at $x=0$, namely $P(x,t)|_{x=0^{-}}= \kappa P(x,t)|_{x=0^{+}}$, where $\kappa$ was assumed to be equal to $\kappa=1$ \cite{CGS05} or $\kappa=const.$ \cite{Kosztolowicz}. For normal diffusion in samples $x<0$ and $x>0$ $\kappa$ was derived using a normal random walk theory \cite{Cornell}. A generalization of the boundary conditions for sub-diffusion with unbiased boundary, was considered in \cite{Morph,Zoia} (see also \cite{MBK00,Silbey,Chechkin,MK00}).

Solution of equation (\ref{anomaleq}) in Laplace space reads 
\begin{eqnarray}
\label{anomal}
\tilde{P}(x,s) = \tilde{C}^{+}(s) \frac{s^{\alpha^{+}/2-1} \exp \left( - \frac{x \; s^{\alpha^{+}/2}}{\sqrt{K^{+}}} \right) }{2 \sqrt{K^{+}}} \theta(x) + \nonumber \\
+ \tilde{C}^{-}(s) \frac{s^{\alpha^{-}/2-1} \exp \left( - \frac{|x| s^{\alpha^{-}/2}}{\sqrt{K^{-}}} \right) }{2 \sqrt{K^{-}}} \left[ 1 - \theta(x) \right],
\end{eqnarray}
where $\tilde{C}^{-}(s)$ and $\tilde{C}^{+}(s)$ are functions soon to be determined. Here $P(x,0)=\delta(x)$ is used for the initial condition.  The conservation of probability: $\int_{-\infty}^{+\infty} dx \tilde{P}(x,s)=1/s$, gives $\tilde{C}^{+}(s)+\tilde{C}^{-}(s)=2$. Using equation (\ref{anomal}) we find  
\begin{equation}
\label{bcanomal2}
s^{-\alpha^{-}+1} K^{-} \frac{\partial \tilde{P}(x,s)}{\partial x}|_{x=0^{-}} - s^{-\alpha^{+}+1} K^{+} \frac{\partial \tilde{P}(x,s)}{\partial x}|_{x=0^{+}} = 1.
\end{equation}
For $\alpha^{-} < 1$ and $\alpha^{+} < 1$, Equation (\ref{bcanomal2}) represents the continuity of fractional probability flow at the boundary
\begin{equation}
\label{J}
J^{+}(x=0^{+},t) - J^{-}(x=0^{-},t) = \delta(t).
\end{equation}
The fractional probability flow in this case is the generalization of the usual definition \cite{MBK99}, for example for $x<0$
\begin{equation}
J^{-}(x,t) = - K^{-} \; _{0}D_{t}^{1-\alpha^{-}} \frac{\partial P(x,t)}{\partial x}.
\end{equation}
Therefore the fractional equation equation (\ref{anomaleq}) can be written as 
\begin{equation}
\partial P(x,t)/\partial t + \partial J^{-}/\partial x = 0,
\end{equation}
for $x<0$ and similarly for $x>0$. Thus, equation (\ref{bcanomal2}) is nothing else but the Laplace transform of the condition of continuity of the fractional probability current at the boundary $J^{-}(x=0^{-},t)=J^{+}(x=0^{+},t)$.

To derive the second boundary 
condition we calculate the first moment, 
$\left< \tilde{x}(s) \right> = \int_{-\infty}^{+\infty} dx \; x \; \tilde{P}(x,s)$ (see Appendix C for the alternative derivation of a second boundary condition). Using equation (\ref{anomal})
\begin{equation}
\label{MPDF}
\left< \tilde{x}(s) \right> = \frac{1}{2s} \left( \sqrt{K^{+}} \tilde{C}^{+}(s) s^{-\frac{\alpha^{+}}{2}} - \sqrt{K^{-}} \tilde{C}^{-}(s) s^{-\frac{\alpha^{-}}{2}} \right).
\end{equation}
\begin{figure}[t!]
  \centering
  \includegraphics[width=.65\textwidth]{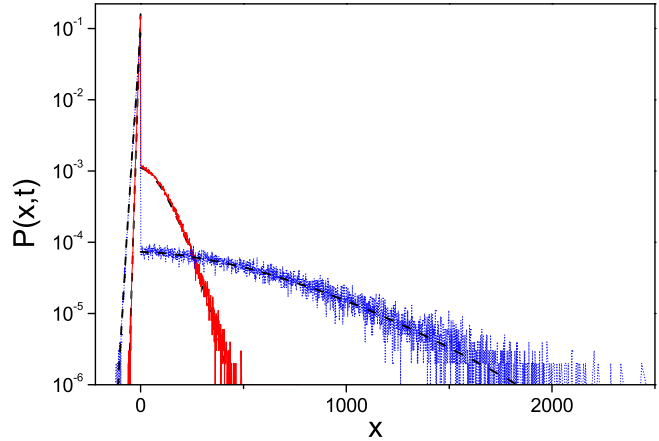}
  \caption{PDF of particle's position $P(x,t)$ calculated 
numerically by CTRW model (dotted lines) perfectly agrees with analytical theory 
(dashed lines) found inverting equations (\ref{anomal}) and (\ref{C}) to the time domain.
The figure illustrates that the majority of the particles are found in the slow
domain $x<0$, however the tail of the PDF extends deeply into the
fast domain. Thus even if eventually all the particles will be accumulated in
the slow domain on the left,  $\langle x \rangle$ will be located in the fast domain on the right. 
We used  $\alpha^{-}=0.3$, $K^{-}=0.385$, $\alpha^{+}=0.8$, $K^{+}=0.108$ and $q^{+}=0.7$ at time $t=10^6$ (solid line) and $t=10^8$ (dotted line). For $t=10^8$ almost all particles are in $x<0$, $\mathcal{P}^{-}\simeq0.93$, but the drift is positive $\left< x \right>=14.2$.}
  \label{FIG2}
\end{figure}
We require that equation (\ref{MPDF}) be equal to the Laplace transform of
equation (\ref{xmean3}), which was calculated from the CTRW model. For $\alpha^{-}<\alpha^{+}$, equations (\ref{MPDF}) and (\ref{xmean3}) yields when $s \rightarrow 0$ 
\begin{equation}
\label{C}
\tilde{C}^{+}(s) \sim \frac{2q^{+}}{q^{-}}\sqrt{\frac{K^{-}}{K^{+}}} s^{\frac{\alpha^{+}-\alpha^{-}}{2}}, \; \; \tilde{C}^{-}(s) = 2 - \tilde{C}^{+}(s).
\end{equation}
Using equations (\ref{C}) and (\ref{anomaleq}), we derive the second boundary condition
\begin{equation}
\label{bcanomal1}
q^{+} K^{-} s^{-\alpha^{-}} \tilde{P}(x,s)|_{x=0^{-}} = q^{-} K^{+} s^{-\alpha^{+}} \tilde{P}(x,s)|_{x=0^{+}}.
\end{equation}
Equation (\ref{bcanomal1}) shows that generally the PDF at the boundary is not continuous, similar to the normal diffusion case \cite{Cornell}. Such a jump in the PDF on the origin is shown in figure \ref{FIG2}. Equation (\ref{bcanomal1}) also shows that the scaling of the solution is more complex than in time-fractional diffusion equation with one diffusion exponent. Using equations (\ref{anomal}) and (\ref{C}) we find the scaling of the solution for $\alpha^{-}<\alpha^{+}$
\begin{equation}
\label{anomaleqscaled}
G_{\alpha^{-},\alpha^{+}}(\xi) = 
\cases{t^{\alpha^{-}/2} \; P(x,t), &for $x<0$ \\
t^{\alpha^{+}-\alpha^{-}/2} \; P(x,t),&for $x>0$, \\}
\end{equation}
and
\begin{equation}
\label{xsi_sub}
\xi = 
\cases{|x|/t^{\alpha^{-}/2}, &for $x<0$ \\
x/t^{\alpha^{+}/2},&for $x>0$. \\}
\end{equation}
Numerically calculated scaled PDF is shown in figure \ref{FIG_scaled_subpdf} and exhibits data collapse.
\begin{figure}[t!]
  \centering
  \includegraphics[width=.65\textwidth]{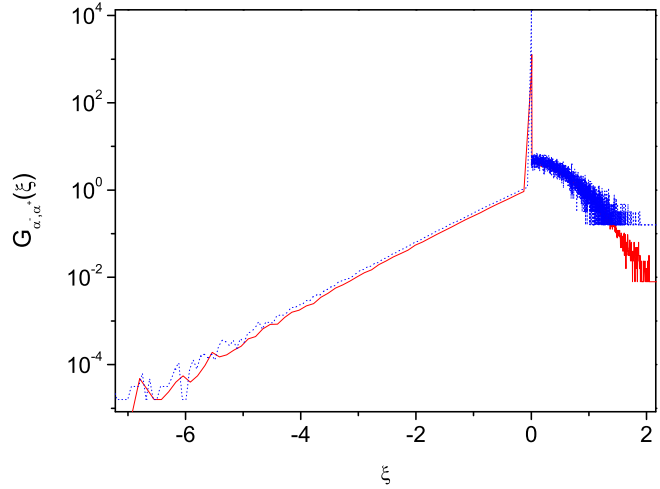}
  \caption{Same as in figure \ref{FIG2} presented in scaled form given by equations (\ref{anomaleqscaled}, \ref{xsi_sub}).}
  \label{FIG_scaled_subpdf}
\end{figure}

It is straightforward to calculate moments using the analytical expression for the propagator equations (\ref{anomal}) and (\ref{C}). For the mean we obtain 
\begin{equation}
\label{xs}
\left< \tilde{x}(s) \right> = \frac{(q^{+} - q^{-}) \sqrt{K^{-} K^{+}}}{s \left( q^{+} \sqrt{K^{-}} s^{\alpha^{+}/2} + q^{-} \sqrt{K^{+}} s^{\alpha^{-}/2} \right)}.
\end{equation}
Expanding equation (\ref{xs}) in $s$ we get the result, which coincides with one calculated from the CTRW model (see equation (\ref{xmean2})). For the second moment we get the following expression
\small
\begin{equation}
\label{x2}
\left< \tilde{x}^2(s) \right> = \frac{2\sqrt{K^{-} K^{+}} \left( q^{-} \sqrt{K^{-}} s^{-\alpha^{-}/2} + q^{+} \sqrt{K^{+}} s^{-\alpha^{+}/2} \right)}{s\left(q^{-} \sqrt{K^{+}} s^{\alpha^{-}/2} + q^{+} \sqrt{K^{-}} s^{\alpha^{+}/2}\right)},
\end{equation}
\normalsize
which in the long time limit $t \rightarrow \infty$ ($s \rightarrow 0$) gives for $\alpha^{-}<\alpha^{+}$
\begin{equation}
\label{x2long}
\left< x^2(t) \right> \sim \frac{q^{+}}{q^{-}} \; \frac{2 \sqrt{K^{-} K^{+}} t^{(\alpha^{-}+\alpha^{+})/2}}{\Gamma(1+(\alpha^{-}+\alpha^{+})/2)}.
\end{equation}
This result is in agreement with numerical simulations of the CTRW model (see figure \ref{FIG}). 

\subsection{Drift against the flow and flow without the drift}

Consider the situation where 
 $\alpha^{-} <\alpha^{+}$, so we call the domain $x<0 $ the ``slow" medium.
From equation (\ref{avfrac2}), the probability to be in the slow medium in the long time limit is
\begin{equation}
\label{p+}
\mathcal{P}^{-}(t) \rightarrow 1, \; \; \; t \rightarrow \infty.
\end{equation} 
We also calculated the mean drift in this case (see equation (\ref{xmean3})).
Thus, as mentioned, independently of the details of the model all particles flow into the slower medium, which absorbs them in the long time limit. However, at the same time if $q^{+} > q^{-}$, the drift  $\langle x \rangle>0 $ is positive and increasing with time. Namely, $\langle x \rangle$ is located in the ``fast" medium even though all particles eventually accumulate in the slow medium. As mentioned, while the dynamics in the faster domain $x>0$ is clearly important (since $\langle x \rangle$ may be in that domain) the mean $\langle x \rangle$ does not depend on the diffusion constant $K^{+}$ of that medium, neither on the anomalous diffusion exponent $\alpha^{+}$. 

An explanation of this paradox is as follows: Although the region with smaller $\alpha$ will accumulate more and more particles in the long time limit, at finite time there will be always some particles in the opposite region where $\alpha$ is larger. As shown in figure \ref{FIG2}, these particles are moving more freely and travel far away from the interface which will compensate the accumulation of particles in the region with smaller $\alpha$. In other words while $\mathcal{P}^{+} = 1 - \mathcal{P}^{-} = \int_{0}^{\infty} P(x,t) dx \rightarrow 0$, which naively implies $\lim_{t \rightarrow \infty} P(x,t) = 0$ for $x>0$ (since $P(x,t)\ge 0$), still $\int_{0}^{\infty} x \; P(x,t) dx$ does not approach zero.

\section{Coupled Sub and Superdiffusive systems}

  Now we consider a composition of subdiffusive material in one region (for example $x<0$) and a material with superdiffusion in the other region ($x>0$). Subdiffusion is modeled by the CTRW on a lattice. As before, a particle has the probability $1/2$ to jump to one of its nearest neighbors. Waiting times on each lattice point are independent identically distributed random variables with a common PDF $\psi(\tau) \propto \tau^{-1-\alpha}$ as $\tau \rightarrow \infty$ with $0<\alpha<1$. Thus, in our composite system a particle starting on the origin will jump say to the right  after waiting a random time drawn from the distribution $\psi(\tau)$, and performs a super-diffusive walk in $x>0$ until it returns and crosses the boundary $x=0$. Then, a particle performs CTRW in $x<0$ until it returns the boundary and so on. At the boundary we consider equal probabilities to go left or right $q^{+}=q^{-}=1/2$.

For superdiffusion we consider the L{\'e}vy walk model, which corresponds to the spatiotemporally coupled version of the CTRW \cite{ZK93,ZKB93}. The waiting time and jump length PDFs are no longer decoupled but appear as $\psi(x,t)=\lambda(x)p(t|x)$. We consider the coupling in the form $p(t|x)=\frac{1}{2}\delta(|x|-v t)$, where $v$ is a velocity and the PDF of jump length $\lambda(x)\propto \sigma^{\gamma}/|x|^{1+\gamma}$ with $0 < \gamma \le 2$. In what follows we consider $v=1$. Since the velocity $v$ is finite it penalizes long jumps such that the overall process attains a finite variance (in contrast to infinite variance of L{\'e}vy flights $\left< x^2(t) \right>=\infty$) \cite{ZK93,ZKB93}. For a L\'evy walk (i.e. without coupling to a subdiffusive system) 
\begin{equation}
\label{x2_LW}
\left< x^2(t) \right> \propto 
\cases{t^{3-\gamma}, &for $1< \gamma \le 2$, \\
t^2, &for $0<\gamma \le 1$. \\}
\end{equation}
For $\gamma = 2$ the L{\'e}vy walk converges to Gaussian process with $\left< x^2(t) \right> \propto t$. 

\begin{figure}[t!]
  \centering
  \includegraphics[width=.85\textwidth]{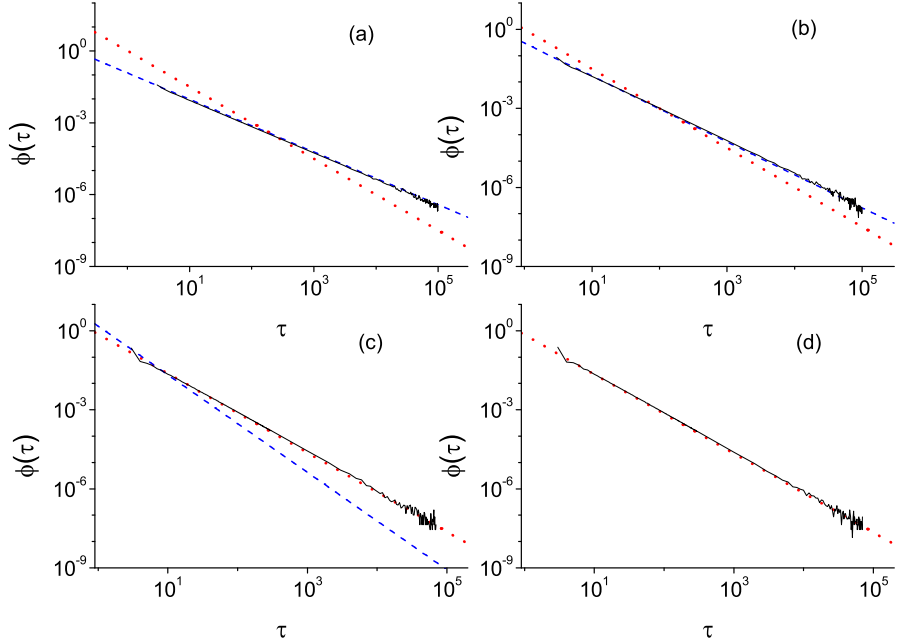}
  \caption{PDF of first passage time $\phi(\tau)$ calculated numerically for the Levy walk on a semi axis for (a) $\gamma=0.2$, (b) $\gamma=0.5$, (c) $\gamma=1.7$, (d) $\gamma=2$. Blue dashed lines are proportional to $\propto \tau^{-1-\gamma/2}$ and red dotted lines are proportional to $\propto \tau^{-3/2}$. Simulations agree with our theoretical prediction equation (\ref{phi_LW}).}
  \label{FIG_subsuper}
\end{figure}

Our aim is to calculate the occupation fractions, the PDF and the drift for the introduced coupled sub and superdiffusive systems. For this we need to know the first passage time (FPT) density of the L\'evy walks. Let us consider the FPT distribution for the L\'evy walks on a semi axis. Since for $\gamma \rightarrow 2$ the L{\'e}vy walk converges to Gaussian process, the PDF of the FPTs for the L{\'e}vy walk with $\gamma=2$ should be $\phi(\tau) \propto \tau^{-3/2}$ \cite{Redner}. For $1<\gamma \le 2$ we find that the PDF of the FPTs for the L{\'e}vy walk to be independent of $\gamma$ while for $0<\gamma < 1$ the PDF of the FPTs depends on $\gamma$ 
\begin{equation}
\label{phi_LW}
\phi(\tau) \propto 
\cases{\tau^{-3/2}, &for $1< \gamma \le 2$, \\
\tau^{-1-\gamma/2}, &for $0<\gamma \le 1$. \\}
\end{equation}
These results can be deduced from the exponents the FPT densities of the subdiffusive CTRW. Coupling a CTRW system with a L\'evy walk system we expect that occupation fractions in both systems will attain finite values only if $\gamma=\alpha$. For that to happen FPT in both systems must behave similarly. Hence for a L\'evy walk with $0<\gamma<1$ the first passage time PDF is $\phi(\tau) \propto \tau^{-1-\gamma/2}$ in the superdiffusive system corresponding to $\phi(\tau) \propto \tau^{-1-\alpha/2}$ in the subdiffusive system. For $1<\gamma<2$ we expect $\phi(\tau) \propto \tau^{-3/2}$ since in that regime the coupled system exhibits normal behavior ($1<\gamma\le 2$ implies normal first passage times). Numerically calculated FPT densities for the L{\'e}vy walks are shown in figure \ref{FIG_subsuper} and they are compatible with equation (\ref{phi_LW}).

\subsection{Occupation fractions}
 
Using the FPT density we can now calculate the distribution of occupation times in the coupled sub and superdiffusive systems. For this we use the method described in section 2.2 (see also Appendix B). The PDF of occupation times is determined by the first passage time PDFs in $x>0$ and $x<0$. For the CTRW in the material $x<0$ the first passage PDF is given by $\phi^{-}(\tau) \propto \tau^{-(1+\alpha/2)}$ with $0<\alpha \le 1$. For Levy walk in the material $x>0$ the first passage PDF is as we just have shown $\phi^{+}(\tau) \propto \tau^{-3/2}$ if $1<\gamma \le 2$ and $\phi^{+}(\tau) \propto \tau^{-(1+\gamma/2)}$ if $0<\gamma < 1$. So, depending on the values of $\alpha$ and $\gamma$ three cases can be classified: Case (I) $1<\gamma \le 2$ and $0<\alpha<1$, so $\alpha<\gamma$. In this case the average occupation fraction in the material $x<0$ behaves as $\mathcal{P}^{-}(t) \propto 1 - t^{-(1-\alpha)/2} \rightarrow 1$ as $t \rightarrow \infty$ and in the long time limit almost all particles will be in the material $x<0$. Accordingly, since $\mathcal{P}^{+}(t)+\mathcal{P}^{-}(t)=1$ 
\begin{equation}
\label{Pt+}
\mathcal{P}^{+}(t) \propto t^{-(1-\alpha)/2} \rightarrow 0, \; t \rightarrow \infty. 
\end{equation}
For $0<\gamma \le 1$ there are two cases: case (II) $0 < \alpha < \gamma \le 1$ and case (III) $0 < \gamma < \alpha \le 1$. When $\alpha<\gamma$ the average occupation fractions in $x<0$ and $x>0$ behave as $\mathcal{P}^{-}(t) \propto 1 - t^{-(\gamma-\alpha)/2} \rightarrow 1$ and $\mathcal{P}^{+}(t) \propto t^{-(\gamma-\alpha)/2} \rightarrow 0$ as $t \rightarrow \infty$. For $\alpha>\gamma$, $\mathcal{P}^{-}(t) \propto t^{-(\alpha-\gamma)/2} \rightarrow 0$ and correspondingly $\mathcal{P}^{+}(t) \propto 1 - t^{-(\alpha-\gamma)/2} \rightarrow 1$ as $t \rightarrow \infty$. Collecting results we write 
\begin{equation}
\label{P-}
\mathcal{P}^{-}(t) \sim 
\cases{1, &(I): $1< \gamma \le 2, \; 0<\alpha \le 1$, \\
0, &(II): $0<\gamma<\alpha \le 1$, \\
1, &(III): $0<\alpha<\gamma \le 1$, \\}
\end{equation}
and $\mathcal{P}^{-}(t)+\mathcal{P}^{+}(t)=1$. This result is very natural, wherever we find the largest sticking time the occupation fraction in that system will be $1$. Only when $\alpha=\gamma$, $\mathcal{P}^{-}$ and $\mathcal{P}^{+}$ are not trivial in the long time limit.

\subsection{Scaling of the PDF}

Using the average occupation fractions we can find the scaling of the particles density. First we consider case (I), namely $1<\gamma \le 2$ and $0<\alpha<1$. 
For the coupled sub and superdiffusive system we are looking for the PDF in $x<0$ in the form (see equation (\ref{anomal}))
\begin{equation}
\label{Pxm}
\tilde{P}(x,s) \propto \tilde{Y}^{-}(s) \frac{s^{\alpha/2-1} \exp \left( - \frac{|x| s^{\alpha/2}}{\sqrt{K_{\alpha}}} \right) }{\sqrt{K_{\alpha}}}, \; \; x<0, 
\end{equation}
where $K_{\alpha}$ is the fractional subdiffusion coefficient. Integrating equation (\ref{Pxm}) we find the average occupation fraction in $x<0$, $\tilde{\mathcal{P}}^{-}(s)=\int_{-\infty}^{0} dx \tilde{P}(x,s) \propto \tilde{Y}^{-}(s)/s$. However, as we have just shown, for case (I) the average occupation fraction $\mathcal{P}^{-}(t) \rightarrow 1$ as $t \rightarrow \infty$ or $\tilde{\mathcal{P}}^{-}(s) \propto 1/s$ as $s \rightarrow 0$, which yields $\tilde{Y}^{-}(s) \propto 1$.
\begin{figure}[t!]
  \centering
  \includegraphics[width=.65\textwidth]{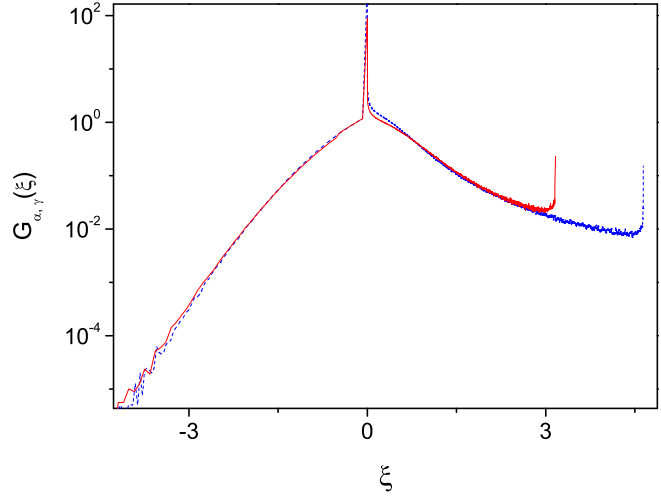}
  \caption{Scaled PDF $G_{\alpha,\gamma}(\xi)$ for the coupled sub and superdiffusive system defined in (\ref{G_I}) calculated numerically at $t=10^3$ and $t=10^4$ (solid and dashed lines) with $\alpha=0.7$ and $\gamma=1.2$. The scaling variable $\xi$ is defined in equation (\ref{xsi_I}). The peaks are a result of ballistic paths, i.e. path with no turn overs.}
  \label{FIG_subsuper_spdf}
\end{figure}
For $x>0$, similar to equation (\ref{Pxm}), we are looking for PDF, which in Fourier-Laplace space has the form
\begin{equation}
\label{Pks}
\hat{\tilde{P}}(k,s)=\tilde{Y}^{+}(s) \; \hat{\tilde{W}}_{\gamma}(k,s). 
\end{equation}
For L\'evy walks with $1<\gamma \le 2$ the center part of the PDF has the form of L\'evy density \cite{ZK93,ZKB93}
\begin{equation}
\label{LWs}
W_{\gamma}(x,t) \sim \frac{1}{(K_{\gamma} t)^{1/\gamma}} l_{\gamma}\left( \frac{x}{(K_{\gamma} t)^{1/\gamma}} \right),
\end{equation}
where 
\begin{equation}
\label{LWs1}
l_{\gamma}(\zeta)=\frac{1}{2 \pi}\int_{-\infty}^{+\infty} e^{- i q \zeta} l_{\gamma}(q) dq, \; \; \zeta=\frac{x}{(K_{\gamma} t)^{1/\gamma}},
\end{equation}
with the characteristic function ($K_{\gamma}$ is some constant)
\begin{equation}
\label{LWs2}
l_{\gamma}(q) = \exp (-K_{\gamma}|q|^{\gamma}),
\end{equation}
provided that the density was initially localized at $x = 0$, and exhibits a sharp cutoff marked by the ballistic peaks at $|x| = v t$ \cite{DKU03} (clearly seen in figure \ref{FIG_subsuper_spdf}).

Taking $k = 0$, or integrating the PDF equation (\ref{Pks}) in $x$ from zero to infinity, we find the average occupation fraction in $x>0$, $\tilde{\mathcal{P}}^{+}(s)=\tilde{Y}^{+}(s)/s$. On the other hand, using equation (\ref{Pt+}) the average occupation fraction in the material $x>0$ in Laplace space behaves as $\tilde{\mathcal{P}}^{+}(s) \propto s^{-1+(1-\alpha)/2}$ for $s \rightarrow 0$. Comparing two expressions we find 
\begin{equation}
\label{Cs}
\tilde{Y}^{+}(s) \propto s^{(1-\alpha)/2}, \; \; s \rightarrow 0, \; \; \; \; \; 
Y^{+}(t) \propto t^{-1-(1-\alpha)/2}, \; \; t \rightarrow \infty.
\end{equation}
Inverting the Fourier-Laplace transforms of equation (\ref{Pks}) and using equations (\ref{LWs}), (\ref{Cs}), the PDF for $x>0$ reads
\begin{equation}
\label{Pxt}
P(x,t) \propto \int_{0}^{t} d \tau  \frac{(t - \tau)^{-1-(1-\alpha)/2}}{\tau^{1/\gamma}} \; l_{\gamma}\left(\frac{x}{(K_{\gamma} \tau)^{1/\gamma}}\right), \; \; x>0. 
\end{equation}

 From equations (\ref{Pxt}), (\ref{Pxm}) it follows that the PDF of particle position for coupled sub-superdiffusive systems with $1<\gamma \le 2$ and $0<\alpha<1$ possesses the scaling form 
\begin{equation}
\label{G_I}
G_{\alpha,\gamma}(\xi) \propto 
\cases{t^{\alpha/2} P(x,t), &for $x<0$ \\
t^{1/\gamma+1/2-\alpha/2} P(x,t),&for $x>0$, \\}
\end{equation}
and
\begin{equation}
\label{xsi_I}
\xi = 
\cases{|x|/t^{\alpha/2}, &for $x<0$ \\
x/t^{1/\gamma},&for $x>0$. \\}
\end{equation}
To reconfirm this result notice that using the scaling of the PDF, the occupation fraction in $x>0$ is
\begin{equation}
\mathcal{P}^{+}(t) = \int_{0}^{\infty} dx \; P(x,t) \sim t^{-1/\gamma-1/2+\alpha/2} \int_{0}^{\infty} dx \; G_{\alpha, \gamma}\left(\frac{x}{t^{1/\gamma}}\right) \sim 
\end{equation}
\[
\sim t^{-1/2+\alpha/2} \int_{0}^{\infty} d\xi \; G_{\alpha, \gamma}(\xi) \propto t^{-1/2+\alpha/2},
\]
which is what we have found from the FPT analysis (see equation (\ref{Pt+})). Numerically calculated scaled PDF is shown in figure \ref{FIG_subsuper_spdf} for $\alpha=0.5$ and $\gamma=1.75$.
Similarly, for the case (II) $0<\alpha<\gamma<1$ 
\begin{equation}
\label{G_II}
G_{\alpha,\gamma}(\xi) \propto 
\cases{t^{\alpha/2} P(x,t), &for $x<0$ \\
t^{1+(\gamma-\alpha)/2} P(x,t),&for $x>0$, \\}
\end{equation}
and
\begin{equation}
\label{xsi_II}
\xi = 
\cases{|x|/t^{\alpha/2}, &for $x<0$ \\
x/t,&for $x>0$. \\}
\end{equation}
\begin{figure}[t!]
  \centering
  \includegraphics[width=.65\textwidth]{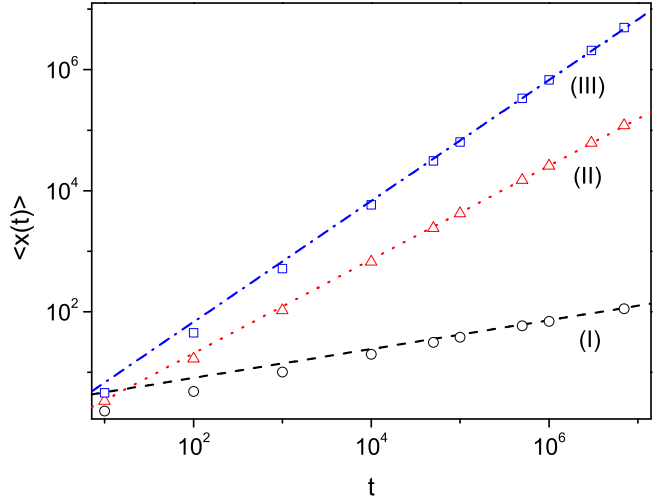}
  \caption{Simulation of the drift $\left< x(t) \right>$ for coupled sub-superdiffusive systems with $q^{+}=q^{-}$ (symbols) are favorably compared with our theory equation (\ref{x_subsuper}) (dashed, dotted and dashed-dotted lines). (I) $0<\alpha<1$, $1<\gamma\le2$: $\alpha=0.3$, $\gamma=1.7$; (II) $0<\alpha<\gamma<1$: $\alpha=0.35$, $\gamma=0.8$; (III) $0<\gamma<\alpha<1$: $\alpha=0.7$, $\gamma=0.3$.}
  \label{FIG_subsuper_mean}
\end{figure}
For the case (III) $0<\gamma<\alpha<1$ the scaling of the PDF for $x>0$ is determined by the ballistic regime of L\'evy walks $\left< |x| \right> \propto t$ \cite{ZK93,ZKB93}
\begin{equation}
\label{G_2_3}
G_{\alpha,\gamma}(\xi) \propto 
\cases{t^{\alpha-\gamma/2} P(x,t), &for $x<0$ \\
t \; P(x,t),&for $x>0$, \\}
\end{equation}
and
\begin{equation}
\label{xsi_2_3}
\xi = 
\cases{|x|/t^{\alpha/2}, &for $x<0$ \\
x/t,&for $x>0$. \\}
\end{equation}

\subsection{Drift in coupled sub-superdiffusive system}

Using the scaling form of the PDF it is easy to estimate the sign and the time dependence of the the mean position of the packet, which is initially at $x=0$. As it follows from equations (\ref{xsi_I}), (\ref{xsi_II}) and (\ref{xsi_2_3}), L\'evy walks always spread further than subdiffusive trajectories: In all cases for $x<0$, $|x| \propto t^{\alpha/2}$ while for $x>0$, $x \propto t^{1/\gamma+1/2-\alpha/2}$ for the case (I) and $x \propto t$ for cases (II) and (III). Therefore, the sign of the mean is always positive, that is the drift is directed to the L\'evy walk independently of $q^{+}/q^{-}$.

Now we calculate the time dependence of the drift 
\begin{equation}
\label{xt}
\left< x (t)\right> = \int_{-\infty}^{\infty} dx \; x \; P(x,t) = \int_{-\infty}^{0} dx \; x \; P(x,t) + \int_{0}^{\infty} dx \; x \; P(x,t).
\end{equation}
For the case (I) $0<\alpha<1$, $1<\gamma\le2$, using the scaling form of the PDF equations (\ref{G_I}, \ref{xsi_I}), we find 
\begin{equation}
\label{ss_x_m}
\int_{-\infty}^{0} dx \; x \; P(x,t) \sim t^{-\alpha/2} \int_{-\infty}^{0} dx \; x \; G_{\alpha, \gamma}\left( \frac{|x|}{t^{\alpha/2}} \right) \sim 
\end{equation}
\[
\sim t^{\alpha/2} \int_{-\infty}^{0} d\xi \; \xi \; G_{\alpha, \gamma}\left( \xi \right) \sim t^{\alpha/2},
\]
\begin{equation}
\label{ss_x_s}
\int_{0}^{\infty} dx \; x \; \tilde{P}(x,t) \sim t^{-1/\gamma-1/2+\alpha/2} \int_{0}^{\infty} dx \; x \; G_{\alpha, \gamma}\left( \frac{x}{t^{1/\gamma}} \right) \sim 
\end{equation}
\[
\sim t^{1/\gamma - 1/2 + \alpha/2} \int_{0}^{\infty} d\xi \; \xi \; G_{\alpha, \gamma}\left( \xi \right) \sim t^{1/\gamma - 1/2 + \alpha/2}.
\]
For $1<\gamma \le 2$ and $0 < \alpha < 1$ the exponent in equation (\ref{ss_x_s}) is always greater than the exponent in (\ref{ss_x_m}), $1/\gamma - 1/2 + \alpha/2 > \alpha/2$, and therefor for the case (I) 
\begin{equation}
\left< x(t) \right> \propto t^{1/\gamma - 1/2 + \alpha/2} > 0, \; t \rightarrow \infty.
\end{equation}
Using equations (\ref{G_II}),(\ref{xsi_II}),(\ref{G_2_3}),(\ref{xsi_2_3}) we find the drift for the cases (II), (III). Summarizing results we have
\begin{equation}
\label{x_subsuper}
\left< x(t) \right> \propto 
\cases{t^{1/\gamma - 1/2 + \alpha/2}, &(I): $1< \gamma \le 2, \; 0<\alpha \le 1$, \\
t^{1 -\gamma/2 + \alpha/2}, &(II): $0<\alpha<\gamma\le 1$, \\
t, &(III): $0<\gamma<\alpha\le1$. \\}
\end{equation}
Figure \ref{FIG_subsuper_mean} demonstrates a good agreement between the theory and numerical simulations for all three cases. Results for the occupation fractions and the drift suggest that for the case (I) and (III) ($0<\alpha<1$, $1<\gamma<2$ and $0<\alpha<\gamma<1$) the drift is opposite to the flow, which for these cases is directed to the subdiffusive part, $\mathcal{P}^{-} \rightarrow 1$ as $t \rightarrow \infty$ (see equation (\ref{P-})).

\section*{Summary}

We investigated systems consisting of two materials with sub or superdiffusive properties and a boundary between them. In coupled subdiffusive system particles flow to the slower medium, while the direction of the averaged drift is determined by symmetry breaking at the boundary, $q_L \ne q_R$ in our model. This leads to interesting phenomena unique to subdiffusion: (i) under certain conditions all particles are found in one sample (e.g. $\mathcal{P}^{-} \rightarrow 1$), but the drift is oppositely directed ($\left< x \right> > 0$), (ii) the drift does not depend on properties of the fast medium, namely even if $\left< x(t) \right> > 0$, the anomalous diffusion exponent $\alpha^{+}$ and $K^{+}$ in $x>0$ do not control $\left< x(t) \right>$ (under certain conditions). We find similar behavior for the diffusion in  quenched trap model and two dimensional comb structure, which points out to a broader generality of our results (see \cite{KB2010}). We argue that such a behavior is a general feature of subdiffusion in disordered systems. For coupled sub-superdiffusive systems we find a net drift, which is always directed to the superdiffusive material independently of the asymmetry at the boundary, while the direction of the particles flow depends on the relation between sub and superdiffusion exponents. These phenomena are explained by the competition of the diffusion processes which are slower or faster than normal spreading.  

\section*{Acknowledgments}

This work is supported by the Israel Science Foundation. We thank D. Kessler, S. Burov and S. Carmi for helpful remarks.

\appendix

\section{Calculation of $\left< n_z(t) \right>$} 

Here we calculate the drift for the system of two coupled subdiffusive materials. For this we need to calculate the average number of returns to the origin, $\left< n_z(t) \right>$, which is determined in the following way: We define a three state process $\xi(t)=0$ (state $0$) if the particle is on the origin, $\xi(t)=+1$ (state $+$) if the particle is in $x>0$ and $\xi(t)=-1$ (state $-$) if the particle is in $x<0$ (see figure \ref{FIG_3CTRW}). In the long time limit the number of visits to the origin is independent of $R_0$ since the average waiting times in state $+$ and $-$ are infinite. The waiting times in state $+$ are the first passage times, from $x=a$ to $x=0$, and similarly the waiting times in state $-$ are the first passage time from $-a$ to $0$. These first passage times in the continuum limit were obtained previously \cite{Barkai01} and they are one sided L\'evy distributions whose Laplace transform is 
\begin{equation}
\label{fpt}
\tilde{\phi}^{-}(s) = e^{- a s^{\alpha^{-}/2}/\sqrt{K^{-}}}, \; \; \; \; \tilde{\phi}^{+}(s) = e^{- a s^{\alpha^{+}/2}/\sqrt{K^{+}}},
\end{equation}
which implies $\phi^{-}(t) \propto t^{-(1+\alpha^{-}/2)}$ and similarly for $\phi^{+}(t)$. For $\alpha^{-}=\alpha^{+}=1$ we get well known distribution of the first passage time of a Brownian motion in half space \cite{Redner}. The Laplace transform of the probability to have exactly $n_z$ transitions is given by \cite{Feller} 
\begin{equation}
\label{P}
\tilde{P}_{n_z}(s) = \frac{1-\tilde{\bar{\phi}}(s)}{s} \tilde{\bar{\phi}}^{n_z}(s), 
\end{equation}
where $\tilde{\bar{\phi}}(s) = q^{-} \tilde{\phi}^{-}(s) + q^{+} \tilde{\phi}^{+}(s)$. Using equation (\ref{P}), the average number of transitions to zero is given by
\begin{equation}
\label{nav}
\left< \tilde{n}_z(s) \right>  = \frac{\tilde{\bar{\phi}}(s)}{s \left( 1 - \tilde{\bar{\phi}}(s)\right)}. 
\end{equation}

The small $s$ expansion yields the long $t$ behavior of $\left< n_z(t) \right>$. Using equations (\ref{fpt}, \ref{nav})
\begin{equation}
\label{nmean2}
\left< \tilde{n}_z(s) \right> \sim 
\cases{\frac{\sqrt{K^{+}}}{a q^{+}} \; s^{-(1+\alpha^{+}/2)}, &for $\alpha^{-} > \alpha^{+}$ \\
\frac{\sqrt{K^{-}}}{a q^{-}} \; s^{-(1+\alpha^{-}/2)},&for $\alpha^{-} < \alpha^{+}$ \\
\left[ s \left( \frac{q^{-} a s^{\alpha^{-}/2}}{\sqrt{K^{-}}} + \frac{q^{+} a s^{\alpha^{+}/2}}{\sqrt{K^{+}}} \right) \right]^{-1},&for $\alpha^{-}=\alpha^{+}.$ \\}
\end{equation}
Inverting equation (\ref{nmean2}) to the time domain and using equation (\ref{xmean}) we find the drift in the long time limit. 


\section{CTRW: Occupation fractions}
\begin{figure}[t!]
  \centering
  \includegraphics[width=.55\textwidth]{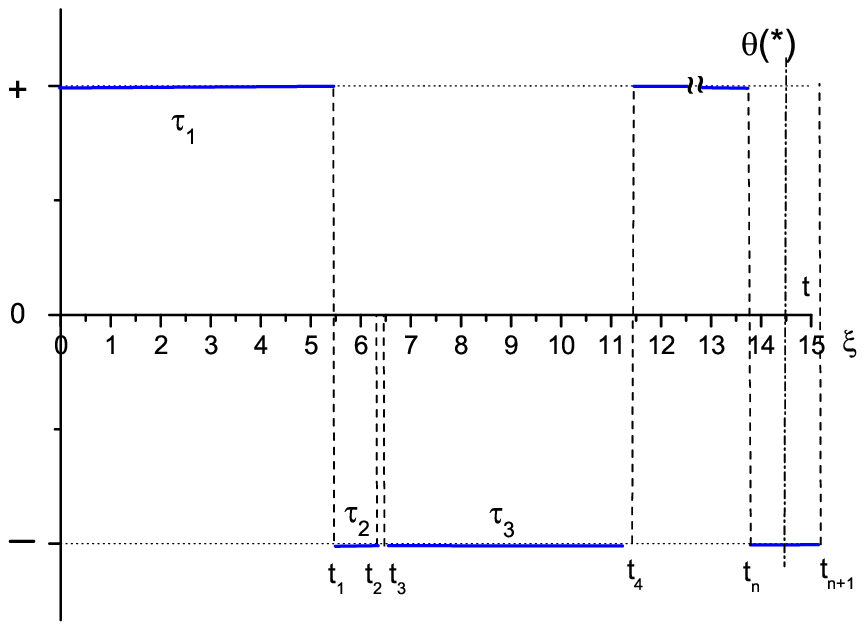}
  \caption{The three state CTRW model constructed from the random walk illustrated in figure \ref{FIG_RW}.}
  \label{FIG_3CTRW}
\end{figure}
Here we calculate the PDF of the occupation fraction in the state $-$ (calculation for the state $+$ is similar) for the CTRW. For that aim we use the three state process defined in Appendix A. The total time a particle was in the state $-$ after $n$ steps can be written as (see figure \ref{FIG_3CTRW}) 
\begin{equation}
\label{Tplus}
t^{-}=\sum_{i=1}^{n} \tau_i \theta_i + (t - t_n) \theta(*),
\end{equation}
where $\theta_i$ is a random variable with the PDF
\begin{equation}
\label{ThetaPDF}
p(\theta) = q^{+} \delta (\theta) + q^{-} \delta (\theta-1),
\end{equation}
that is $\theta_i=0$ when the position of the particle is $x>0$ or in state $+$ and $\theta_i=1$ for $x<0$ (state $-$).
In equation (\ref{Tplus}) $\theta (*)$ denotes the last step and $t_n=\sum_{i=1}^{n} \tau_i = \sum_{i=1}^{n} \tau_i \theta_i + \sum_{i=1}^{n} \tau_i |\theta_i - 1|$. 
The PDF of the occupation time $t^{-}$ after time $t$ and $n$ steps is defined as 
\begin{equation}
\label{pdfplus}
f_{t,n}(t^{-}) = \left< \delta \left( t^{-} - \sum_{i=1}^{n} \tau_i \theta_i - (t - t_n) \theta(*) \right) \; 
I(t_n \le t \le t_{n+1}) \right>,
\end{equation}
where
\begin{equation}
\label{I}
I(t_n \le t \le t_{n+1}) =\cases{1&if condition in parentheses is true,\\
0&otherwise.\\}
\end{equation}

Now we consider the double Laplace transform of equation (\ref{pdfplus})
\begin{eqnarray}
\label{pdfflus}
\tilde{f}_{s,n}(u) = \int_{0}^{\infty} dt \; e^{-st} \int_{0}^{\infty} dt^{-} \; e^{-u t^{-}} f_{n,t^{-}} = \nonumber\\
=\left< \int_{0}^{\infty} dt \; e^{-s t}I(t_n \le t \le t_{n+1}) e^{ -u \sum_{i=1}^{n} \tau_i \theta_i - u (t - t_n) \theta(*)} \; \right>. 
\end{eqnarray}
Averaging over the last step $\theta(*)$, we get
\begin{eqnarray}
\tilde{f}_{s,n}(u) = q^{+} \left< \frac{e^{-s t_n} - e^{-s t_{n+1}}}{s} \; e^{-u \sum_{i=1}^{n} \tau_i \theta_i} \right> + \nonumber\\
+ q^{-} \left< \frac{e^{-(s+u) t_n} - e^{-(s+u) t_{n+1}}}{s+u} \; e^{-u \sum_{i=1}^{n} \tau_i \theta_i + u \sum_{i=1}^{n} \tau_i} \right>.
\end{eqnarray}
Averaging now over $\theta_i$ and summing over all jumps we get the PDF of $t^{-}$ 
\begin{eqnarray}
\label{ap_ofrac}
\tilde{f}_{s}(u) = \sum_{n=0}^{\infty} f_{s,n}(u) = \nonumber\\
=\frac{1}{1-\left( q^{-} \phi^{-}(u+s) + q^{+} \phi^{+}(s) \right)} \; \left( q^{-} \frac{1-\phi^{-}(u+s)}{u+s} + q^{+} \frac{1-\phi^{+}(s)}{s}\right)
\end{eqnarray}
where $\phi^{\pm}$ are the waiting time PDFs in states $+$ and $-$. Using the long time limit (or small $s$) $\phi^{\pm}$ given by $\tilde{\phi}^{-}(s) \propto 1 - a s^{\alpha^{-}/2}/\sqrt{K^{-}}$ and $\tilde{\phi}^{+}(s) \propto 1 - a s^{\alpha^{+}/2}/\sqrt{K^{+}}$ \cite{Barkai01} as $s \rightarrow 0$ (see equation (\ref{fpt})), we finally obtain equation (\ref{ofrac}). 

For $q^{-}=q^{+}$ and $\alpha^{-}=\alpha^{+}=\alpha$ equation (\ref{ofrac}) is reduced to the Lamperti PDF \cite{GL01,BB06}
\begin{equation}
\label{fs}
\tilde{f}_{s}(u) \sim \frac{ \sqrt{\frac{K^{-}}{K^{+}}} (s+u)^{\alpha/2-1} + s^{\alpha/2-1}}{ \sqrt{\frac{K^{-}}{K^{+}}} (s+u)^{\alpha/2} + s^{\alpha/2}},
\end{equation}
which is a generalization of well-known arcsine law \cite{Feller}. The method of inversion of equation (\ref{fs}) to time domain is given in \cite{GL01}.

\section{Remark on the solution of equation (\ref{anomal})}

We note that the solution of fractional equations (\ref{anomal}) and (\ref{C}) must be used with care. While the PDF of particle's position $P(x,t)$ calculated numerically by CTRW model perfectly agrees with analytical theory including the jump at the boundary (figure \ref{FIG2}) and while this solution gives the correct asymptotic behavior of the occupation fraction $\mathcal{P}^{-} \rightarrow 1$ (when $\alpha^{-}<\alpha^{+}$) (see equation (\ref{avfrac2})), using equations (\ref{anomal}) and (\ref{C}) the occupation fraction $\mathcal{P}^{-}(t)=\int_{-\infty}^{0}dx \; x \; P(x,t)$ in sample $x<0$ within the fractional framework is 
\begin{equation}
\label{Pp}
\mathcal{P}^{-}(t) \propto 1 -  \frac{q^{+}\sqrt{K^{-}}}{q^{-} \sqrt{K^{+}}} \; \frac{t^{\frac{\alpha^{-}-\alpha^{+}}{2}}}{\Gamma \left( 1 + \frac{\alpha^{-}-\alpha^{+}}{2} \right)}.
\end{equation}
Note that while equation (\ref{Pp}) gives correct leading term, the correction term differs from the exact CTRW result (\ref{avfrac2}) (compare the Gamma functions). Figure \ref{FIG_P+} illustrates the difference of two solutions. Thus, fractional equation works in the long time limit and already the first correction to asymptotic solution shows deviation from exact result.

An alternative to (\ref{bcanomal1}) boundary condition can be derived by requiring the equality of occupation fractions calculated by the CTRW model (\ref{Pp}) with the occupation fractions obtained from the fractional equation (\ref{avfrac2})
\begin{equation}
\label{CC}
 \mathcal{G} \; q^{+} \sqrt{K^{-}} s^{\alpha^{+}/2} \tilde{C}^{-}(s)  = q^{-} \sqrt{K^{+}} s^{\alpha^{-}/2} \tilde{C}^{+}(s), 
\end{equation}
where $\mathcal{G}=1+\frac{\alpha^{+}-\alpha^{-}}{2}$. This boundary condition leads to the solution 
\begin{eqnarray}
\label{eq11}
\tilde{C}^{-}(s) = \frac{2}{1+\mathcal{G}\frac{q^{+}}{q^{-}}\sqrt{\frac{K^{-}}{K^{+}}}s^{\frac{\alpha^{+}-\alpha^{-}}{2}}}, 
\; \; \;
\tilde{C}^{+}(s) = \frac{2}{1+\mathcal{G}^{-1}\frac{q^{-}}{q^{+}}\sqrt{\frac{K^{+}}{K^{-}}}s^{\frac{\alpha^{-}-\alpha^{+}}{2}}}.
\end{eqnarray}
Analytical probability density function equations (\ref{anomal}) and (\ref{eq11}) is different compared to equations (\ref{anomal}) and (\ref{C}). As the latter was derived from the equality of occupation fractions calculated by the CTRW model (\ref{Pp}) and the drift obtained with the fractional equation (\ref{avfrac2}), it does not describe the jump of the PDF at the boundary. This solution also gives different results for the moments including the drift.


\end{document}